%% file: draft_v25.tex
\documentclass[aps,prd,twocolumn,superscriptaddress,showpacs,preprintnumbers,amsmath,amssymb]{revtex4-2}
\usepackage[utf8]{inputenc}

\usepackage[dvipsnames]{xcolor}
\usepackage[normalem]{ulem}
\usepackage{comment}
\usepackage{overpic}
\usepackage{multirow}
\usepackage{booktabs} 

\usepackage[colorlinks,linkcolor=red,anchorcolor=blue,citecolor=blue,urlcolor=blue,breaklinks=true]{hyperref}
\newsavebox{\tablebox}
\usepackage{orcidlink}

\usepackage{graphicx}
\usepackage{dcolumn}
\usepackage{bm}
\usepackage{gensymb}
\usepackage{tabularx}
\usepackage{booktabs}
\usepackage{multirow}
\usepackage{lipsum}
\usepackage{subfigure}

\newcommand{\BESIIIorcid}[1]{\href{https://orcid.org/#1}{\hspace*{0.1em}\raisebox{-0.45ex}{\includegraphics[width=1em]{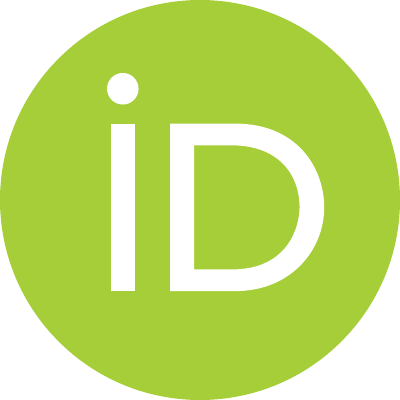}}}}

\begin{document}

\title{Search for the electromagnetic Dalitz decays $\chi_{cJ}\to e^+e^-\phi$}

\input{authorlist_2025-02-08}

\begin{abstract}
Using a data sample of $(2.712 \pm 0.014)\times10^{9}$ $\psi(3686)$ events collected at $\sqrt{s}=3.686$ GeV by the BESIII detector, we search for the rare electromagnetic Dalitz decays $\chi_{cJ}\to e^+e^-\phi~(J=0,\,1,\,2)$ via the radiative transitions $\psi(3686)\to\gamma \chi_{cJ}$. No statistically significant $\chi_{cJ}\to e^+e^-\phi$ signals are observed. The upper limits on the branching fractions of $\chi_{cJ}\to e^+e^-\phi~(J=0,\,1,\,2)$, excluding the $\phi$ resonance to $e^+e^-$ final states, are set to be $2.4\times10^{-7},~6.7\times10^{-7}$ and $4.1\times10^{-7}$ at 90\% confidence level, respectively. This is the first search for the electromagnetic Dalitz transition of P-wave charmonium $\chi_{cJ}$ states to a light vector meson.

\end{abstract}

\maketitle

\section{INTRODUCTION}
Studies of the electromagnetic~(EM) Dalitz decays $\text{M}_1\to\text{M}_2 \gamma^* \to\text{M}_2 l^+ l^- $, M stands for mesons, $l = e~\text{or}~\mu$, play an important role in the understanding of the intrinsic structure of mesons and the interactions between photons and mesons with the EM field~\cite{Landsberg:1985gaz}. In these decays, the lepton pair is generated by the conversion of a virtual photon ($\gamma^*$), and the four-momentum transferred by the virtual photon corresponds to the invariant mass of the lepton pair. Up to now, the EM Dalitz decays of the light unflavored vector mesons $\rho$, $\omega$ and $\phi$ have been intensively studied by the CMD2, SND, NA60 and KLOE experiments~\cite{Fallot:2005liv,CMD-2:2000qla,CMD-2:2000rxk,Achasov:2000ne,NA60:2009una,DiDonato:2010zz}. Moreover, the EM Dalitz decays of the S-wave charmonium state $J/\psi$ have been widely studied by the BESIII experiment, such as $J/\psi\to Pe^+ e^-~(P=\eta^{\prime},\eta,\pi^0)$~\cite{BESIII:2014dax}, $J/\psi\to e^+ e^-\phi$~\cite{BESIII:2019ldo}, $J/\psi\to e^+ e^- \eta$~\cite{BESIII:2018qzg}, $J/\psi\to e^+ e^- \eta'$~\cite{BESIII:2018aao}, $J/\psi\to e^+ e^- X(1835),X(2120),X(2370)$~\cite{BESIII:2021xoh}, $J/\psi\to e^+ e^- \eta(1405)$~\cite{BESIII:2023jyg}. However, studies of analogous transitions of the excited charmonium states are less comprehensive, especially for the P-wave states $\chi_{cJ}$~(throughout this paper, $\chi_{cJ}$ refers to $\chi_{c0},~\chi_{c1}, \text{and}~\chi_{c2}$). Further investigations of $\chi_{cJ}$ Dalitz decays provide valuable insights into the dynamics of P-wave charmonium states, and contribute to a more comprehensive understanding of the EM interaction between photons and charmonium states~\cite{Luchinsky:2017pby}. 

The differential decay width of the EM Dalitz decay $\text{M}_1\to e^+e^- \text{M}_2$ is described by
\begin{eqnarray}
    \frac{d\Gamma}{dq^2} = |F(q^2)|^2\times [\frac{d\Gamma}{dq^2}]_{\rm{point}},
    \label{eq1}
\end{eqnarray}where $[\frac{d\Gamma}{dq^2}]_{\rm{point}}$ is a pointlike assumption that can be calculated with high accuracy by the methods of quantum electrodynamics~\cite{Landsberg:1985gaz}, $q^2$ is the squared four-momentum of the $e^+e^-$ pair, and $F(q^2)$ is the transition form factor~(TFF) used to describe the structure of the EM transition vertex~\cite{Landsberg:1985gaz}. The TFF provides information on the EM structure arising at the vertex of the EM Dalitz decay of mesons, and can be utilized to distinguish between proposed transition mechanisms, such as the dominance of the quark triangle anomalies and quark triangle loops~\cite{Landsberg:1985gaz}. The vector-meson dominance~(VMD) model is the most widely employed model to describe the TFF. With a monopole approximation, the TFF based on the VMD is expressed as
\begin{equation}
\left|F\left(q^{2}\right)\right|=\frac{1}{\left(1-q^{2} / \Lambda^{2}\right)},
\label{eq2}
\end{equation}
where the parameter $\Lambda$ serves as an effective pole mass, subsuming effects from all possible vector resonance poles and scattering terms in the time-like kinematic region~\cite{Landsberg:1985gaz,Fu:2011yy}. The experimental results of the TFF extracted from the study of the EM Dalitz decays of the S-wave charmonium state $J/\psi$~\cite{BESIII:2014dax} are consistent with the predictions of the TFF based on the VMD model~\cite{Fu:2011yy}.

\begin{figure}[htbp]
    \centering
    \subfigure[]{\includegraphics[scale=0.24]{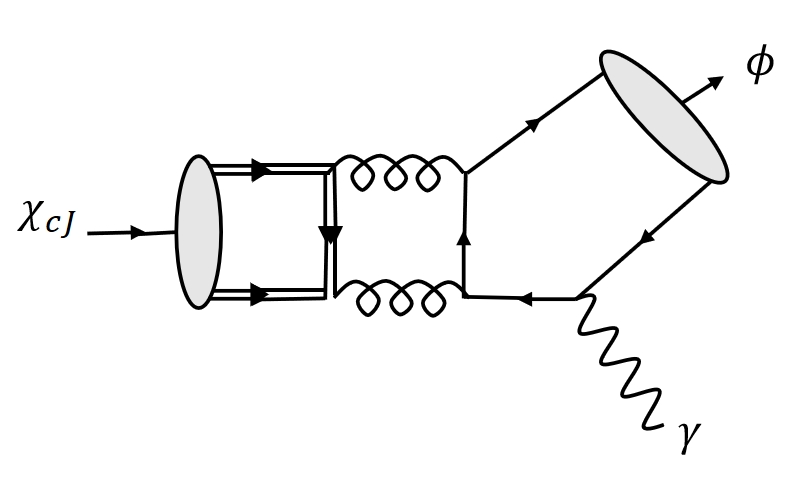}}
    \subfigure[]{\includegraphics[scale=0.24]{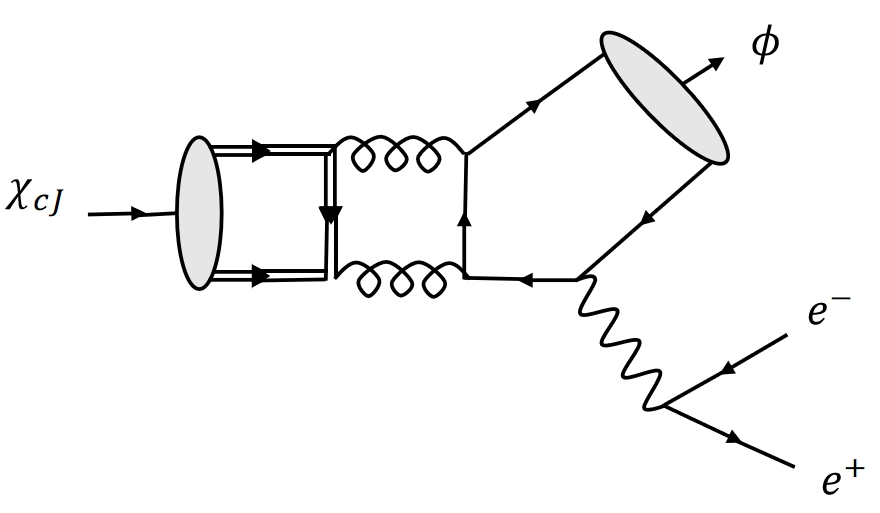}}
    \caption{Feynman diagrams of (a) $\chi_{cJ}\to\gamma\phi$ and (b) $\chi_{cJ}\to e^+e^-\phi$ decays.}
    \label{feynman}
\end{figure}

The charmonium regime lies in the transition region between perturbative and non-perturbative Quantum Chromodynamics~(QCD), and the non-perturbative QCD effects on the transition of charmonium to the final states of light hadrons. The decays of $\chi_{cJ}$ particles are highly sensitive to the behavior of low-energy non-perturbative QCD, making theoretical calculations challenging and heavily reliant on model-based approaches~\cite{Gao:2006bc,Gao:2007fv}. Measurements of the branching fraction $\mathcal{B}(\chi_{cJ}\to e^+e^-\phi)$ offer important experimental insights into charmonium dynamics in the low-energy non-perturbative QCD regime. Additionally, as shown in the Feynman diagrams of Fig.~\ref{feynman}, the $\chi_{cJ}\to \phi e^+e^-$ process has one more EM decay vertex than the $\chi_{cJ}\to\gamma\phi$. The branching fractions are therefore expected to be two orders of magnitude lower than for the corresponding radiative decays $\chi_{cJ}\to\gamma\phi$~\cite{ParticleDataGroup:2024cfk}. A significant deviation from theoretical predictions could indicate hints of new physics, such as the existence of dark photons or other phenomena beyond the Standard Model~\cite{Li:2009wz}.

In this paper, we present the first experimental search for the EM Dalitz decays $\chi_{cJ}\to e^+e^-\phi$ using a data sample of $(2.712 \pm 0.014)\times10^{9}~\psi(3686)$ collected with the BESIII detector~\cite{BESIII:2024lks}.

\section{BESIII DETECTOR AND MONTE CARLO SIMULATION}
The BESIII detector~\cite{ABLIKIM2010345} records symmetric $e^+e^-$ collisions provided by the BEPCII storage ring~\cite{IPAC2016} in the center-of-mass energy ($\sqrt{s}$) range from 1.84 to 4.95~GeV, with a peak luminosity of $1.1 \times 10^{33}\;\text{cm}^{-2}\text{s}^{-1}$ achieved at $\sqrt{s} = 3.773\;\text{GeV}$. BESIII has collected large data samples in this energy region~\cite{BESIII:2020nme,rdtm1,rdtm2}. The cylindrical core of the BESIII detector covers 93\% of the full solid angle and consists of a helium-based multilayer drift chamber~(MDC), a plastic scintillator time-of-flight system~(TOF), and a CsI(Tl) electromagnetic calorimeter~(EMC), which are all enclosed in a superconducting solenoidal magnet providing a 1.0~T magnetic field~(0.9~T in 2012). The solenoid is supported by an octagonal flux-return yoke with resistive plate counter muon identification modules interleaved with steel. The charged-particle momentum resolution at $1~{\rm GeV}/c$ is $0.5\%$, and the ${\rm d}E/{\rm d}x$ resolution is $6\%$ for electrons from Bhabha scattering. The EMC measures photon energies with a resolution of $2.5\%$ ($5\%$) at $1$~GeV in the barrel (end cap) region. The time resolution in the TOF barrel region is 68~ps, while that in the end cap region is 110~ps. The end cap TOF system was upgraded in 2015 using multigap resistive plate chamber technology, providing a time resolution of 60~ps, which benefits 83\% of the data used in this analysis~\cite{rdtm3,rdtm4,Cao:2020ibk}.

Monte Carlo (MC) simulated samples produced with a {\sc geant4}-based~\cite{GEANT4:2002zbu} software package, which includes the geometric description of the BESIII detector~\cite{Huang:2022wuo, Li:2024pox, LANGE2001152, Ya-Jun:2008mrr} and the detector response, are used to determine detection efficiencies and to estimate backgrounds. The inclusive MC sample includes the production of the $\psi(3686)$ resonance incorporated in {\sc kkmc}~\cite{JADACH2000260,PhysRevD.63.113009}. The subsequent decays are modeled with {\sc evtgen}~\cite{Lange:2001uf,Ping:2008zz} using branching fractions either taken from the Particle Data Group~(PDG)~\cite{ParticleDataGroup:2024cfk}, when available, or otherwise estimated with {\sc lundcharm}~\cite{Chen:2000tv,Yang:2014vra}. Final state radiation~(FSR) from charged final state particles is incorporated using the {\sc photos} package~\cite{Barberio:1990ms}. 

The signal MC events of $\chi_{cJ}\to e^+e^-\phi$ are generated using a $q^2$-dependent decay amplitude as described in Ref.~\cite{Landsberg:1985gaz}. The $e^+e^-$ angular distribution is modeled as ($1+\alpha\cdot\cos^2\theta$), where $\theta$ represents the helicity angle in its mother rest frame and $\alpha$ is the angular distribution parameter~\cite{BESIII:2017ung}. For the $\chi_{c0}$ and $\chi_{c2}$ channels, the angular distribution of the former is expected to be flat~\cite{ad8ba2}, and the $\alpha$ of the latter is also assumed to be zero due to the lack of experimental results. The value of $\alpha$ for the $\chi_{c1}$ channel is fixed at 0.565 according to Ref.~\cite{BESIII:2011ysp}. The amplitude is corrected by the TFF given by Eq.~\ref{eq2}, where the $\Lambda$ value is taken as 3.686 GeV/$c^2$~\cite{Fu:2011yy}.

\section{DATA ANALYSIS}

$\chi_{cJ}$ is from $\psi(3686)$ together with a photon, and the decay of interest is $\chi_{cJ}\to e^+e^-\phi$ with $\phi$ reconstructed from a charged kaon pair, so a signal event has 4 charged and 1 neutral tracks. Charged tracks detected in the MDC are required to be within a polar angle ($\theta$) range of $|\rm{cos\theta}|<0.93$, where $\theta$ is defined with respect to the $z$-axis, which is the symmetry axis of the MDC. The distance of closest approach to the interaction point (IP) must be less than 10\,cm along the $z$-axis, and less than 1\,cm in the transverse plane. The number of good charged tracks is required to be equal to four with zero net charge. Particle identification for charged tracks combines measurements of the energy deposited in the MDC ($dE/dx$) and the flight time in the TOF to form the probability values Prob($i$) ($i=e,\pi,K$). A track is considered as an electron when Prob($e$)$>$Prob($\pi$) and Prob($e$)$>$Prob($K$) and as a kaon when Prob($K$)$>$Prob($\pi$) and Prob($K$)$>$Prob($e$). The electron and kaon identification efficiencies are  90\% and 94\%, respectively. The probabilities of misidentifying $h$ as $h^{\prime}$, P($h \to h^{\prime}$), are estimated to be 0.2\% for P($\pi\to e$), 0.2\% for P($K\to e$), 4\% for P($\pi\to K$) and 1\% for P($e\to K$)~\cite{BESIII:2022cmg}. Photon candidates are identified using isolated showers in the EMC. The deposited energy of each shower must be more than 25~MeV in the barrel region ($|\cos \theta|< 0.80$) and more than 50~MeV in the end cap region ($0.86 <|\cos \theta|< 0.92$). To exclude showers that originate from charged tracks, the opening angle subtended by the EMC shower and the position of the closest charged track at the EMC must be greater than 10 degrees as measured from the IP. To suppress electronic noise and showers unrelated to the event, the difference between the EMC time and the event start time is required to be within [0,~700]\,ns. At least one photon is required in each event.

The selected charged tracks are constrained to originate from a common vertex point,  and the $\chi^2$ of the vertex fit is required to be less than 100. A four-constraint (4C) kinematic fit~\cite{Yan:2010zze} is performed by constraining the total four-momentum to that of the initial beams. For events with more than one photon candidate, the combination with the minimum $\chi^2_{\rm{4C}}$ from the 4C kinematic fit is retained. The requirement of $\chi^2_{\rm{4C}}$ is optimized based on the figure-of-merit $\epsilon/(a/2+\sqrt{B})$~\cite{Punzi:2003bu}, where $a$ is the signal significance, $\epsilon$ and $B$ stand for the detection efficiency estimated by the signal MC samples and the number of background events from the inclusive MC samples, respectively. To consider different signal significance assumptions, we performed three separate $\chi^2_{\rm{4C}}$ scans for $a=1,~2,~3$, all of which yielded the same requirement $\chi^2_{\rm{4C}}<40$. Finally, the value of $\chi^2_{\rm{4C}}$ is required to be less than 40. To suppress backgrounds from $\pi^{\pm}/e^{\pm}$ misidentification, the $E/p$ ratio of positron or electron candidates is required to be $0.8<E/p<1.1$ for those entering the EMC, where $E$ refers to the energy deposited in the EMC and $p$ refers to the momentum measured by the MDC. The $e^{\pm}$ candidates not reaching the EMC are all retained. 

Some peaking backgrounds arise from the radiative decays $\chi_{cJ}\to \gamma\phi$ with the photon conversion $\gamma\to e^+e^-$ in the detector material. The conversion mostly occurs at circles of radii around 3.5 cm and 6.5 cm, indicating the positions of the beam pipe and of the inner wall of the MDC, respectively. To effectively reject this background contribution, a $\gamma$-conversion algorithm is implemented to reconstruct the vertex point of the $e^+e^-$ pairs~\cite{Xu2012}, and the distance between the conversion vertex point and the IP in the $xy$ plane, $R_{xy}$, is required to be less than 2 cm. 

\begin{figure*}[!htp]
    \centering
	\subfigure[]{
    \includegraphics[scale=0.29]{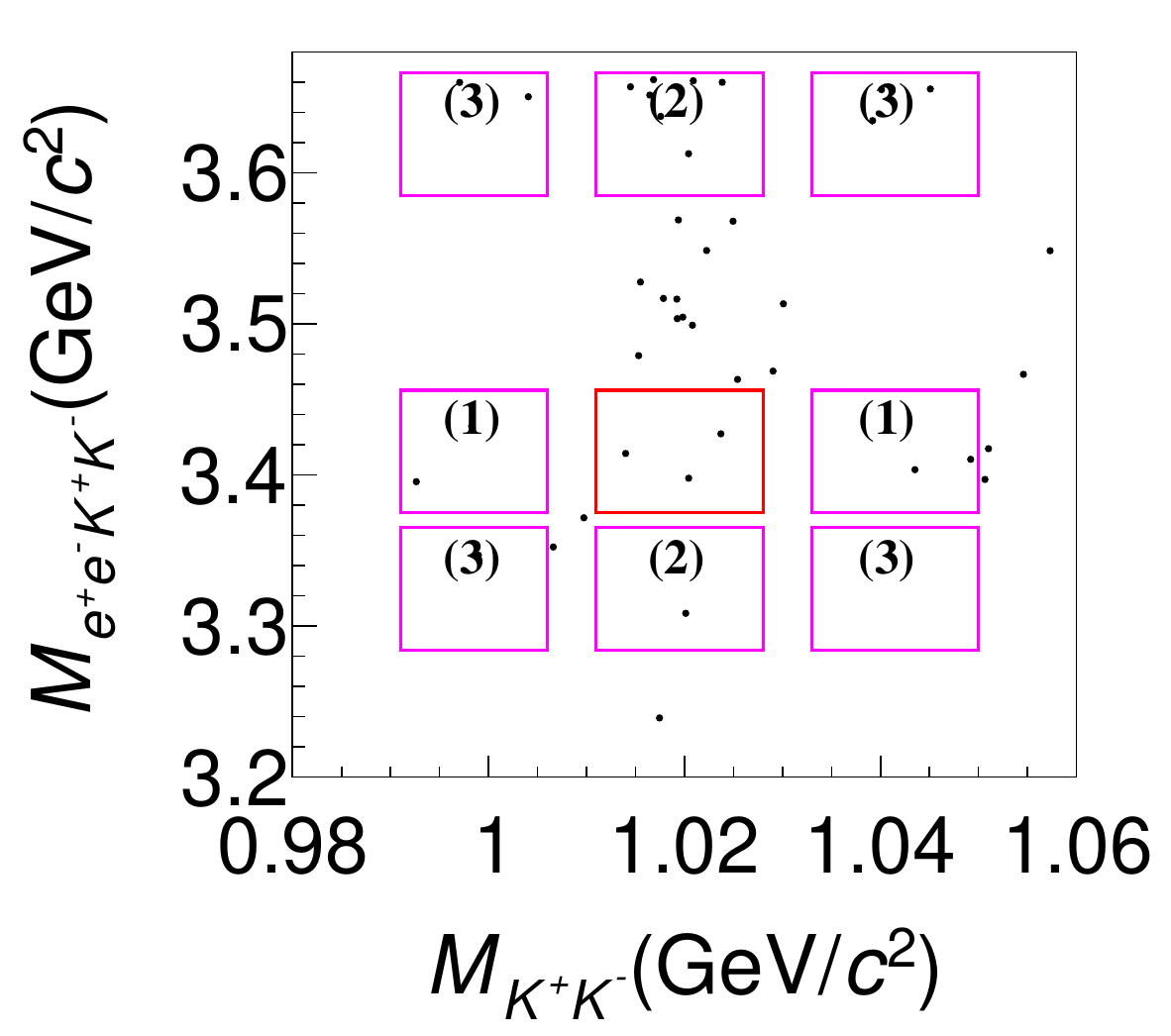}}\subfigure[]{
    \includegraphics[scale=0.29]{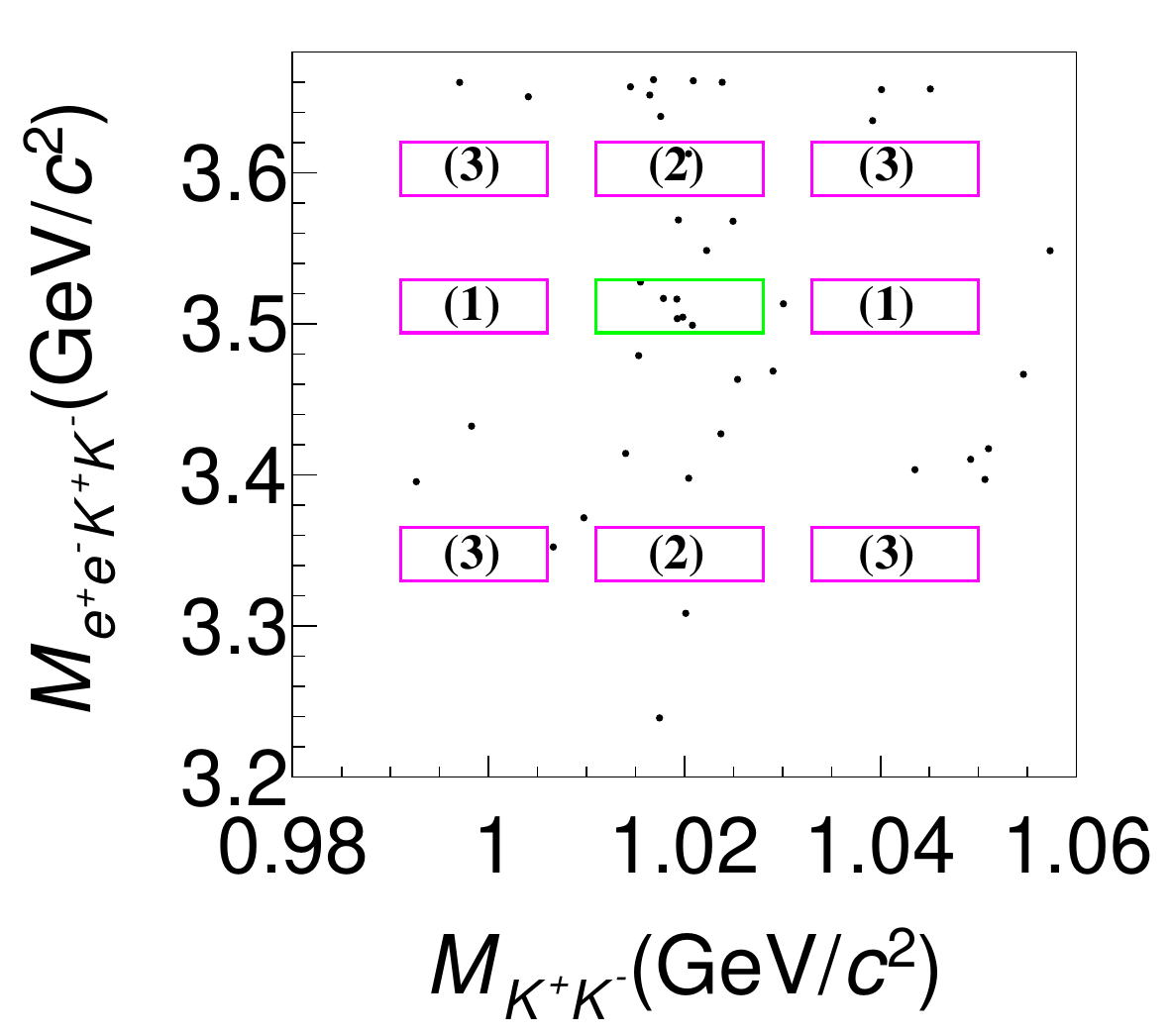}}\subfigure[]{
    \includegraphics[scale=0.29]{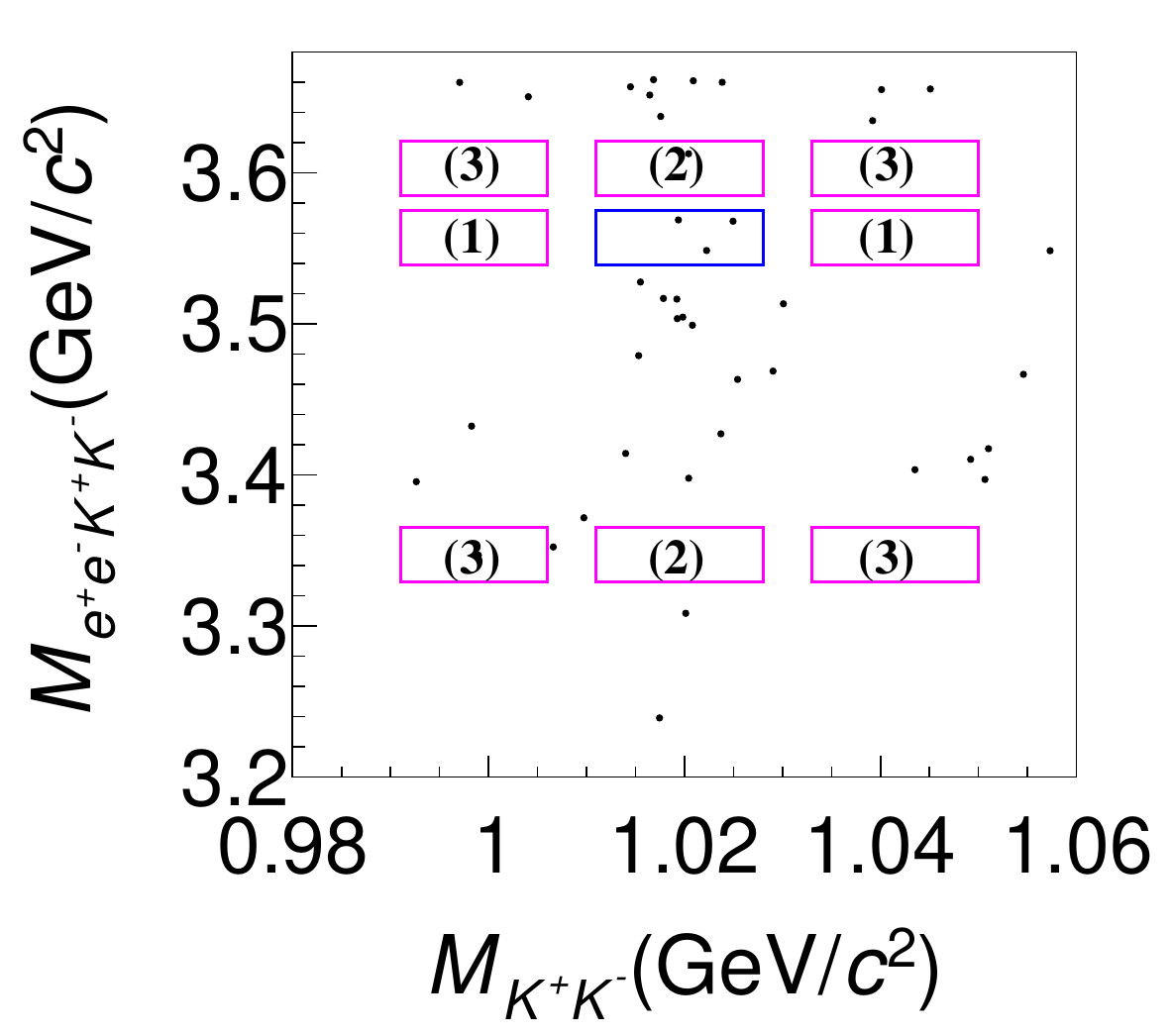}}
    \caption{The distributions of $M_{e^+e^-K^+K^-}$ versus $M_{K^+K^-}$ of the accepted candidate events for (a) $\chi_{c0}$, (b) $\chi_{c1}$, and (c) $\chi_{c2}$ decays into $e^+e^-\phi$ in data, where the red/green/blue boxes are the signal regions and the magenta boxes are the sideband regions. The sideband regions of non-$\phi$ components are indicated as (1), those of non-$\chi_{cJ}$ components as (2), and of neither $\phi$ nor $\chi_{cJ}$ components as (3).}
    \label{2Dsideband}
\end{figure*}

To determine the reject region to suppress backgrounds from $\psi(3686)\to \eta\phi,~\eta\to\gamma e^+e^-,~\phi\to K^+K^-$ and $\psi(3686)\to \gamma\chi_{cJ},~\chi_{cJ}\to\phi\phi,~\phi\to e^+e^-,~\phi\to K^+K^-$, we fit the invariant mass distributions $M_{\gamma e^+e^-}$ and $M_{e^+e^-}$ of the corresponding exclusive MC samples. A double Gaussian function is used to describe the $\eta$ and $\phi$ resonances. The events with $M_{\gamma e^+e^-}$ or $M_{e^+e^-}$ within $5\sigma$ around  $M_{\eta}$ or $M_{\phi}$ are rejected, respectively, where $M_{\eta}$ and $M_{\phi}$ are the mean values of the fitted double Gaussian functions, and $\sigma$ is the corresponding weighted resolution.

The $\phi$ mass window is determined by fitting the $M_{K^+K^-}$ distribution of the signal MC sample with a double Gaussian function to describe the $\phi$ peak. The $\phi$ candidates are selected from $|M_{K^+K^-}-M_{\phi}|<8.5~\text{MeV}/c^2$, corresponding to about $2\sigma$.
The $\chi_{cJ}$ mass windows are determined using the same method adopted for the $\phi$ mass window. The $\chi_{cJ}$ candidates are selected from  $|M_{e^+e^-K^+K^-}-M_{\chi_{c0}}|<40.5~\text{MeV}/c^2$,~$|M_{e^+e^-K^+K^-}-M_{\chi_{c1}}|<17.5~\text{MeV}/c^2$,~$|M_{e^+e^-K^+K^-}-M_{\chi_{c2}}|<18.0~\text{MeV}/c^2$, respectively. The width of the $\chi_{cJ}$ mass windows corresponds to about $3\sigma$. Sideband regions were selected to be symmetric about and non-overlapping with the signal peaks of the $\phi$ and $\chi_{cJ}$~($J=0, 1, 2$) resonances. For $\phi$ meson, the lower and upper sidebands are defined as intervals located $\pm5$ MeV from the boundaries of the signal region, each with a width of $4\sigma$. In terms of invariant mass, this corresponds to the regions (0.99, 1.005)$\cup$(1.033, 1.053) GeV/$c^2$. For the $\chi_{cJ}$ states, the lower sidebands are placed 
10 MeV below the signal region of $\chi_{c0}$ state and the upper sidebands 10 MeV above the signal region of $\chi_{c2}$ state, each with a width of $6\sigma$. The specific mass intervals are (3.284, 3.365)$\cup$(3.585, 3.621) GeV/$c^2$ for the $\chi_{c0}$, (3.330, 3.365)$\cup$(3.585, 3.620) GeV/$c^2$ for the $\chi_{c1}$, and (3.329, 3.365)$\cup$(3.585, 3.621) GeV/$c^2$ for the $\chi_{c2}$. The signal and sideband regions are shown in Fig.~\ref{2Dsideband}.

\begin{figure}[htbp]
    \centering
    \includegraphics[scale=0.4]{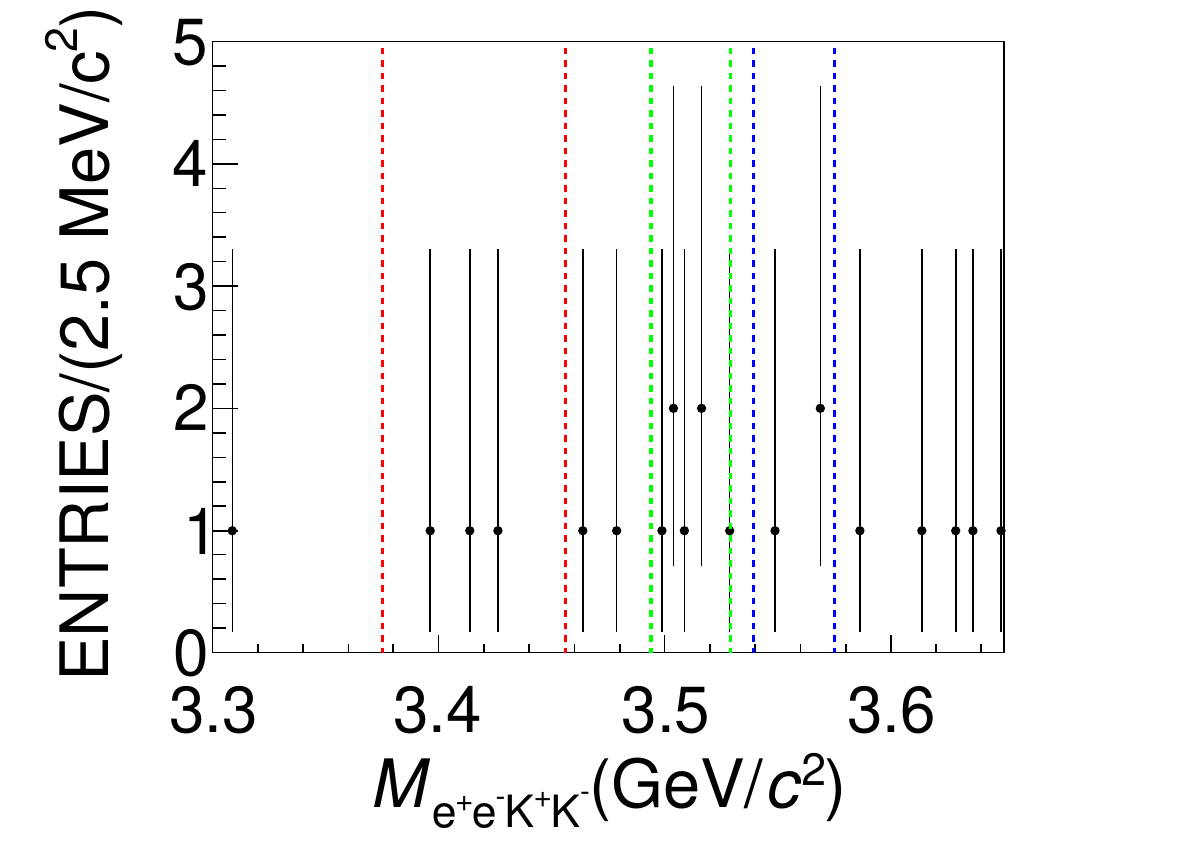}
    \caption{The distribution of $M_{e^+e^-K^+K^-}$. The areas within the red, green, and blue dashed lines represent the $\chi_{c0}$, $\chi_{c1}$, and $\chi_{c2}$ decays into $e^+e^-\phi$, respectively.}
    \label{M_eekk}
\end{figure}

\begin{figure}[htbp]
    \centering
    \includegraphics[scale=0.4]{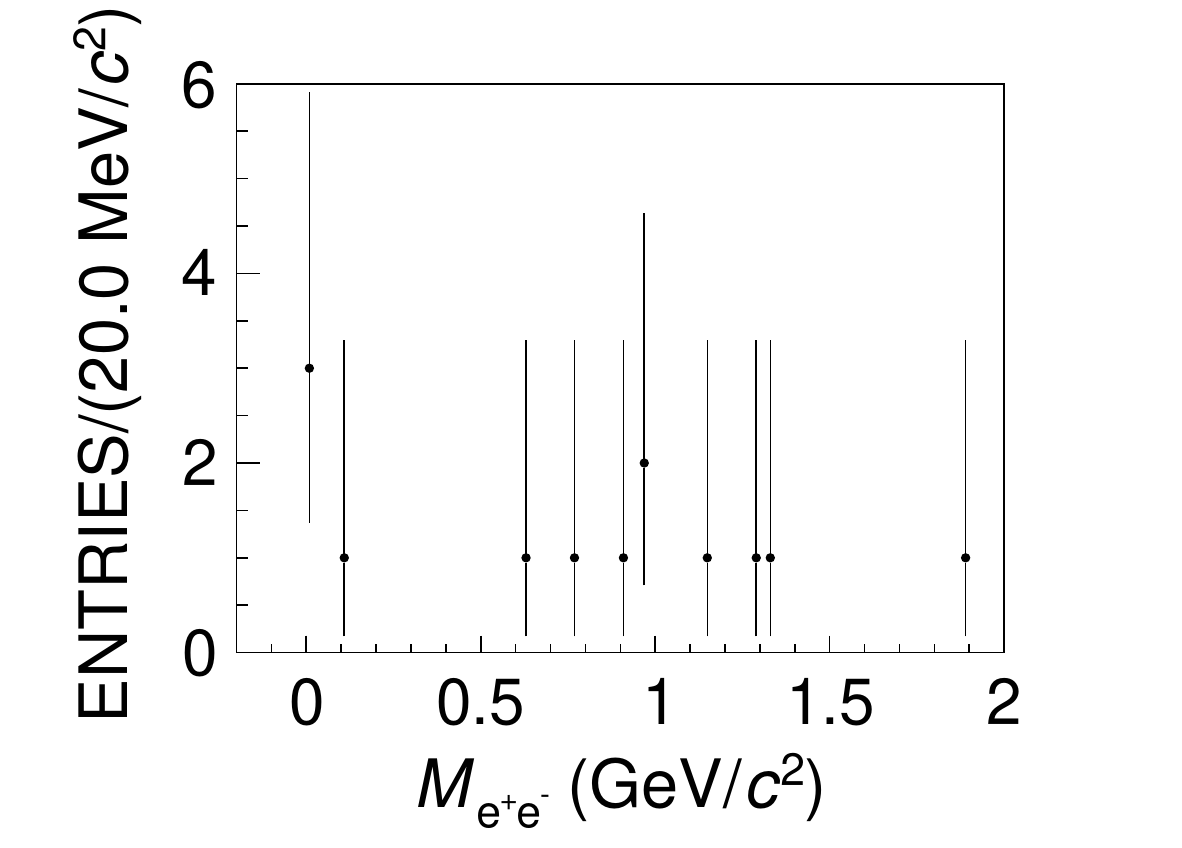}
    \caption{The distribution of $M_{e^+e^-}$.}
    \label{Mee}
\end{figure}

 The distributions of $M_{e^+e^-K^+K^-}$ versus $M_{K^+K^-}$ of the accepted candidate events in data, after applying all the selection criteria, are shown in Fig.~\ref{2Dsideband}, while the distributions of $M_{e^+e^-K^+K^-}$ and $M_{e^+e^-}$ are shown in Fig.~\ref{M_eekk} and Fig.~\ref{Mee}, respectively.

 Since no significant signals were observed in the data, the upper limits on the $\mathcal{B}(\chi_{cJ}\to e^+e^-\phi)$ are given in this analysis. The numbers of events, used to calculate the upper limits of the $\mathcal{B}(\chi_{cJ}\to e^+e^-\phi)$ in Section~\ref{RESULT}, were estimated from the signal and sideband regions. The detection efficiencies were estimated by studying the signal MC samples of $\chi_{c0,1,2}\to e^+e^-\phi$, and are $(12.3\pm0.1)\%,~(13.4\pm0.1)\%,~(12.4\pm0.1)\%$, respectively. 

The comparison of the remaining number of events in the signal region~($N_{\rm obs}$), and the number of background events~($N_{\rm bkg}$), estimated by the sideband method, is listed in Tab.~\ref{compare_nobs_nbkg}. 

\begin{table}[htbp]
  \centering
  \caption{The comparison of $N_{\rm obs}$ and $N_{\rm bkg}$, where the uncertainties are estimated by assuming that the number of events follows a Poisson distribution.}
\begin{tabular}{lcc}
\toprule
\toprule
\makebox[0.15\textwidth]{Decay process} & \makebox[0.15\textwidth][c]{$N_{\rm{obs}}$} & \makebox[0.15\textwidth][c]{$N_{\rm{bkg}}$}\\ \midrule
$\chi_{c0}\to e^+e^-\phi$ & $3$ & $7.1^{+5.1}_{-3.5}$ \\ \midrule
$\chi_{c1}\to e^+e^-\phi$ & $7$ & $0.46^{+3.0}_{-1.0}$ \\ \midrule
$\chi_{c2}\to e^+e^-\phi$ & $3$ & $0.46^{+3.0}_{-1.0}$ \\ 
\bottomrule
\bottomrule
\end{tabular}
\label{compare_nobs_nbkg}
\end{table}%

\section{SYSTEMATIC UNCERTAINTIES}
The sources of systematic uncertainties include the tracking, the PID, the $E/p$ requirement, the photon detection, the 4C kinematic fit, the $\phi$ mass window, the $\chi_{cJ}$ mass window, the $\gamma$-conversion veto, the $\phi\to e^+e^-$ veto, the $\eta\to \gamma e^+e^-$ veto, the angular distribution of $\gamma^{*}$, the TFFs, the sideband region choices, the total number of $\psi(3686)$ events, the $\mathcal{B}(\phi\to K^+K^-)$, and the $\mathcal{B}(\psi(3686)\to \gamma \chi_{cJ})$.

The tracking efficiencies for charged kaons have been studied with the process $J/\psi \to K_S^0K^{\pm}\pi^{\mp}, K_S^0\to\pi^+\pi^-$. The difference in efficiencies between data and MC simulation is 1.0\% per $K^{\pm}$~\cite{BESIII:2011ysp}. The tracking efficiencies for $e^{\pm}$ are obtained with the control sample of radiative Bhabha scattering $e^+e^-\to\gamma e^+e^-$ (including $J/\psi\to \gamma e^+e^-$) at $\sqrt{s} = 3.097~\text{GeV}/c^2 $~\cite{BESIII:2014dax}, and the uncertainty is determined to be 1.0\% per $e^{\pm}$. The total uncertainty due to tracking of four charged tracks is 4.0\%.
The systematic uncertainty from the $e^{\pm}$ PID is investigated by analyzing a mixed control sample of the radiative Bhabha scattering events $e^+e^- \to \gamma e^+e^-$ and the decay $\psi(3686)\to (\gamma_{\rm{FSR}})e^+e^-$ at $\sqrt{s} = 3.686~\text{GeV}$~\cite{BESIII:2022jde}. The systematic uncertainties due to the $E/p$ requirement are estimated in Refs.~\cite{BESIII:2017pez,BESIII:2018iig}  to be 1.0\% per $e^{\pm}$ for PID and 2.0\% per $e^{\pm}$ for $E/p$ requirement. The systematic uncertainty of $K^{\pm}$ PID is studied with the control sample of $e^+e^-\to\pi^+\pi^-J/\psi,~J/\psi\to K^+K^-K^+K^-$~\cite{PhysRevD.109.072015} and is determined to be 1.0\% per $K^{\pm}$.

The uncertainty due to photon detection is 1.0\% per photon based on studies of photon detection efficiencies in $J/\psi \to \pi^0 \pi^+ \pi^-$, $\pi^0 \to \gamma \gamma$~\cite{BESIII:2010ank}. 

For the uncertainty caused by the 4C kinematic fit, the difference is determined by comparing the signal efficiencies before and after the helix correction~\cite{PhysRevD.87.012002}. The uncertainties are determined to be 0.6\%, 0.8\%, and 1.1\% for $\chi_{c0},~\chi_{c1},~\text{and}~\chi_{c2}$ decays, respectively.

To estimate the uncertainties from the $\phi\to e^+e^-$ and $\eta\to \gamma e^+e^-$ veto regions, we use the exclusive MC samples to first extract the $\phi$ and $\eta$ shapes in the $M(e^+e^-)$ and $M(\gamma e^+e^-)$ distributions, respectively. Then, we convolve them with a Gaussian function parameterized by $\Delta\mu$ and $\Delta\sigma$, where $\Delta\mu$ and $\Delta\sigma$ represent the mean value and the width of the resolution difference between data and MC samples. The difference in efficiency estimated by MC samples before and after the smearing is taken as the systematic uncertainty. The systematic uncertainty related to the $\eta\to \gamma e^+e^-$ veto is taken as 0.3\%, while it is negligible for $\phi\to e^+e^-$.

Since the $\phi$ mass window is determined by fitting the M($K^+K^-$) distribution from the signal MC samples, the difference in efficiency caused by the difference of the resolution between data and MC samples is taken as the systematic uncertainties, by applying the same smearing method used for the $\phi\to e^+e^-$ and $\eta\to \gamma e^+e^-$ vetoes.  The systematic uncertainty related to the $\phi$ mass window is taken as 0.7\%.

The systematic uncertainties from the $\chi_{cJ}$ mass windows are estimated by the control sample $\psi(3686)\to \gamma \chi_{cJ}$, $\chi_{cJ}\to \pi^+\pi^-K^+K^-$, by comparing the mass resolution difference between data and MC simulation. The related systematic uncertainties are taken as 0.2\% and 0.1\% for the $\chi_{c0}$ and $\chi_{c2}$ decays, respectively, while the uncertainty for the $\chi_{c1}$ decay is found to be negligible.

The systematic uncertainty due to the $\gamma$-conversion veto, 1.0\%, is estimated by the control sample $J/\psi\to\pi^{+}\pi^{-}\pi^{0},~\pi^0\to\gamma e^+e^-$~\cite{BESIII:2014dax}.

In the MC generation, the virtual photon $\gamma^{*}$ is emitted following the angular distribution 1+$\alpha \cos^2\theta$, where $\theta$ represents the helicity angle in its mother rest frame and $\alpha$ is the angular distribution parameter~\cite{BESIII:2017ung}. The $\alpha$  value for the $\chi_{c0}$ meson is theoretically zero, and does not affect the systematic uncertainty. For $\chi_{c1}$ we vary the $\alpha$ value in the range (0.22, 0.88) according to Ref.~\cite{BESIII:2011ysp}, and in the range (-1, 1) for $\chi_{c2}$. The systematic uncertainties are determined to be 1.0\% and 2.1\% for the $\chi_{c1}$ and $\chi_{c2}$ decays, respectively.

The signal MC events are generated with including the TFF correction. To determine the systematic uncertainties due to the TFFs, we generate the signal events using the TFF from the VMD, with the monopole approximation $F(q^2)=1/(1-q^2/\Lambda^2)$, with $\Lambda = 3.773,~4.040~\rm{GeV}$~\cite{Fu:2011yy}. The maximum differences of signal efficiencies for the $\chi_{c0},~\chi_{c1}$, and $\chi_{c2}$ decays, which are 1.8\%, 1.6\%, and 1.1\%, respectively, are taken as the corresponding systematic uncertainties caused by the TFFs.

The total number of $\psi(3686)$ events has been determined with inclusive $\psi(3686)$ hadronic events, and its uncertainty is assigned as 0.5\%~\cite{BESIII:2024lks}.

In the calculations of the upper limits on the $\mathcal{B}(\chi_{cJ}\to e^+e^-\phi)$, the values of $\mathcal{B}(\psi(3686)\to\gamma \chi_{cJ})$ and $\mathcal{B}(\phi\to K^+K^-)$ are taken from PDG~\cite{ParticleDataGroup:2024cfk}, and their uncertainties are 2.5\%, 3.0\%, and 2.7\% for $\chi_{c0},~\chi_{c1},$ and$\chi_{c2}$ decays, respectively.

Table~\ref{tab:su} provides a summary of the systematic uncertainties affecting the determination of the $\mathcal{B}(\chi_{cJ}\to e^+e^-\phi)$ upper limits. The total systematic uncertainties are determined by combining all individual sources in quadrature.

\begin{table}[!bhtp]
  \centering
  \caption{Summary of systematic uncertainties (\%).}
    \begin{tabular}{lccc}
    \toprule
    \toprule
\makebox[0.25\textwidth]{Source} & \makebox[0.06\textwidth][c]{$\chi_{c0}$} & \makebox[0.06\textwidth][c]{$\chi_{c1}$} & \makebox[0.06\textwidth][c]{$\chi_{c2}$} \\ \midrule
Tracking & 4.0 & 4.0 & 4.0 \\
PID & 4.0 & 4.0 & 4.0 \\
$E/p$ requirement & 4.0 & 4.0 & 4.0 \\
Photon detection & 1.0 & 1.0 & 1.0 \\
Kinematic fit & 0.6 & 0.8 & 1.1 \\
$\eta\to\gamma e^+e^-$ veto & 0.3 & 0.3 & 0.3 \\
$\phi$ mass window & 0.7 & 0.7 & 0.7 \\
$\chi_{cJ}$ mass window & 0.2 & - & 0.1 \\
$\gamma$-conversion veto & 1.0 & 1.0 & 1.0 \\
$\gamma^{*}$ angular distribution& - & 1.0 & 2.1 \\
TFFs & 1.8 & 1.6 & 1.1 \\
Number of $\psi(3686)$ events & 0.5 & 0.5 & 0.5 \\
Intermediate $\mathcal{B}$ & 2.5 & 3.0 & 2.7 \\ \midrule
Total & 7.8 & 8.0 & 8.1 \\
    \bottomrule
    \bottomrule
    \end{tabular}%
  \label{tab:su}%
\end{table}%

\section{RESULTS}
\label{RESULT}
\begin{figure*}[htbp!]
	\centering
	\subfigure[]{
		\includegraphics[scale=0.28]{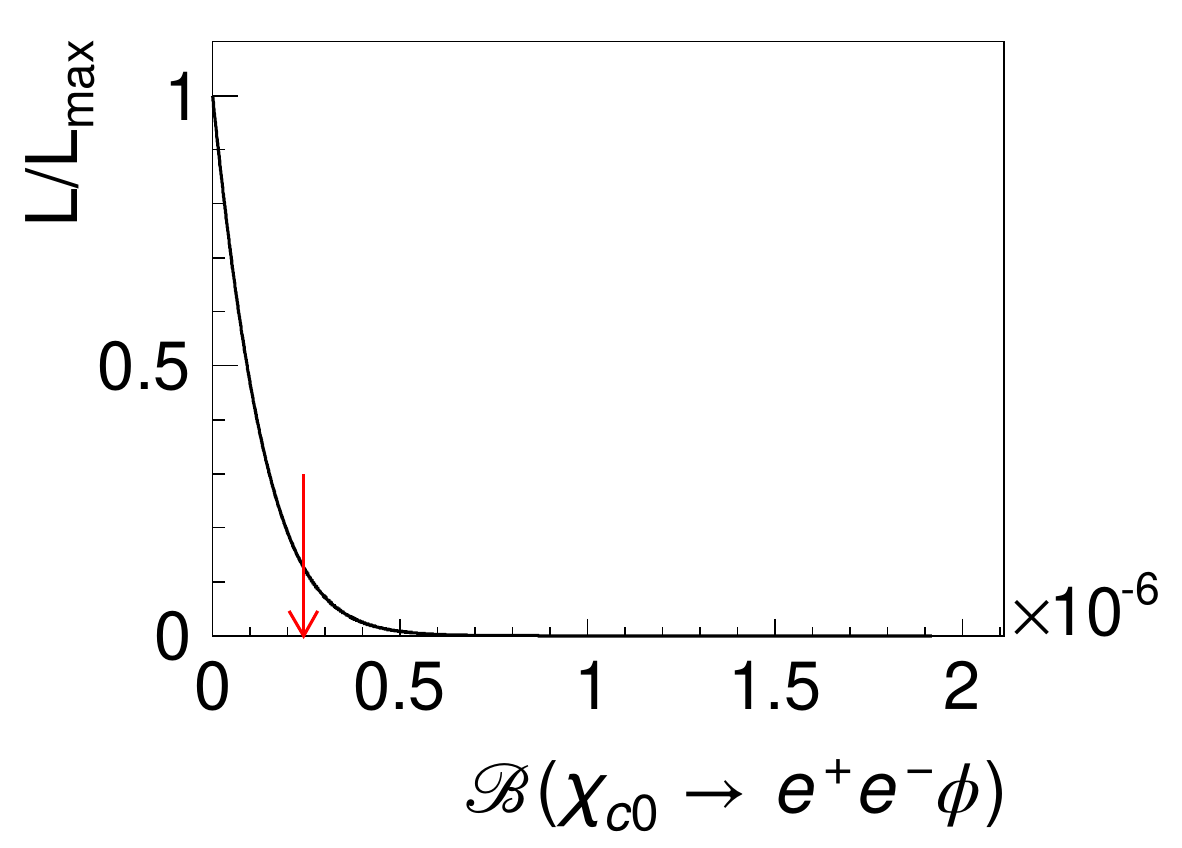}
		\label{}}\subfigure[]{
		\includegraphics[scale=0.28]{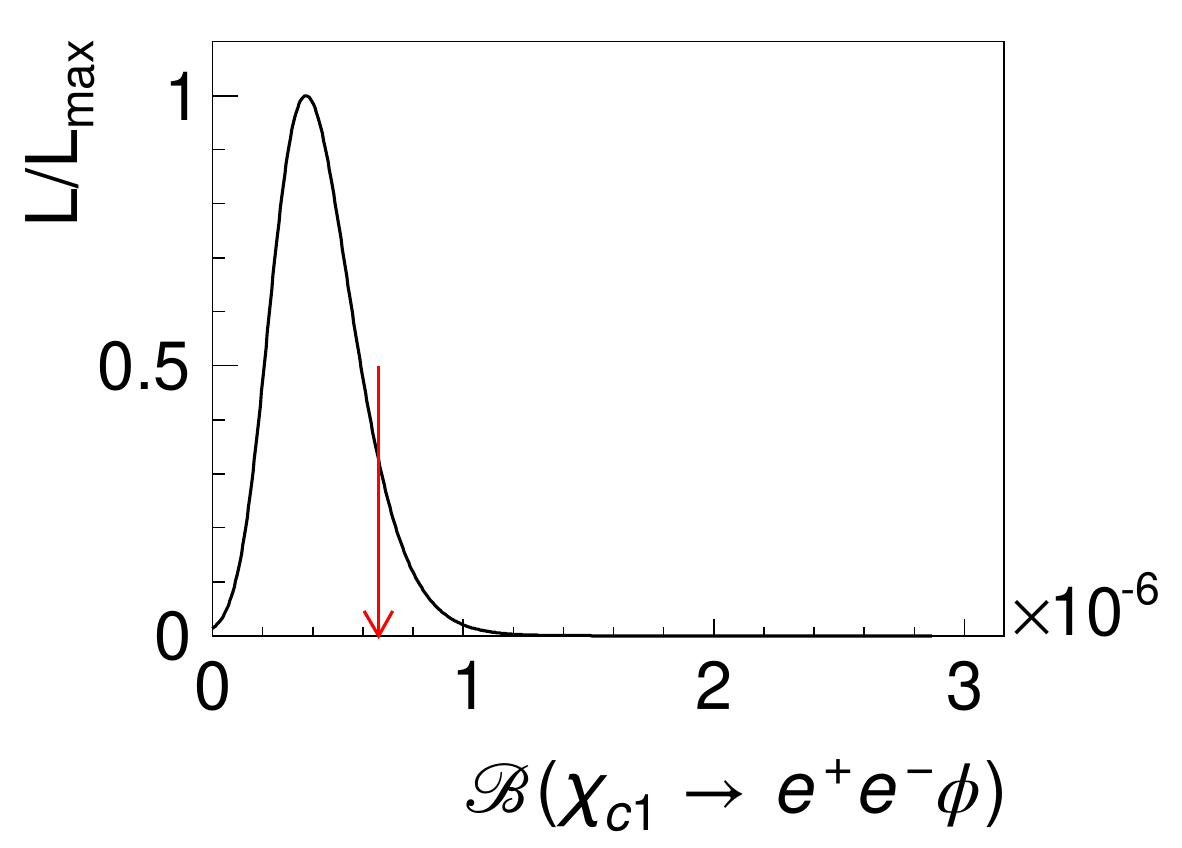}
		\label{}}\subfigure[]{
		\includegraphics[scale=0.28]{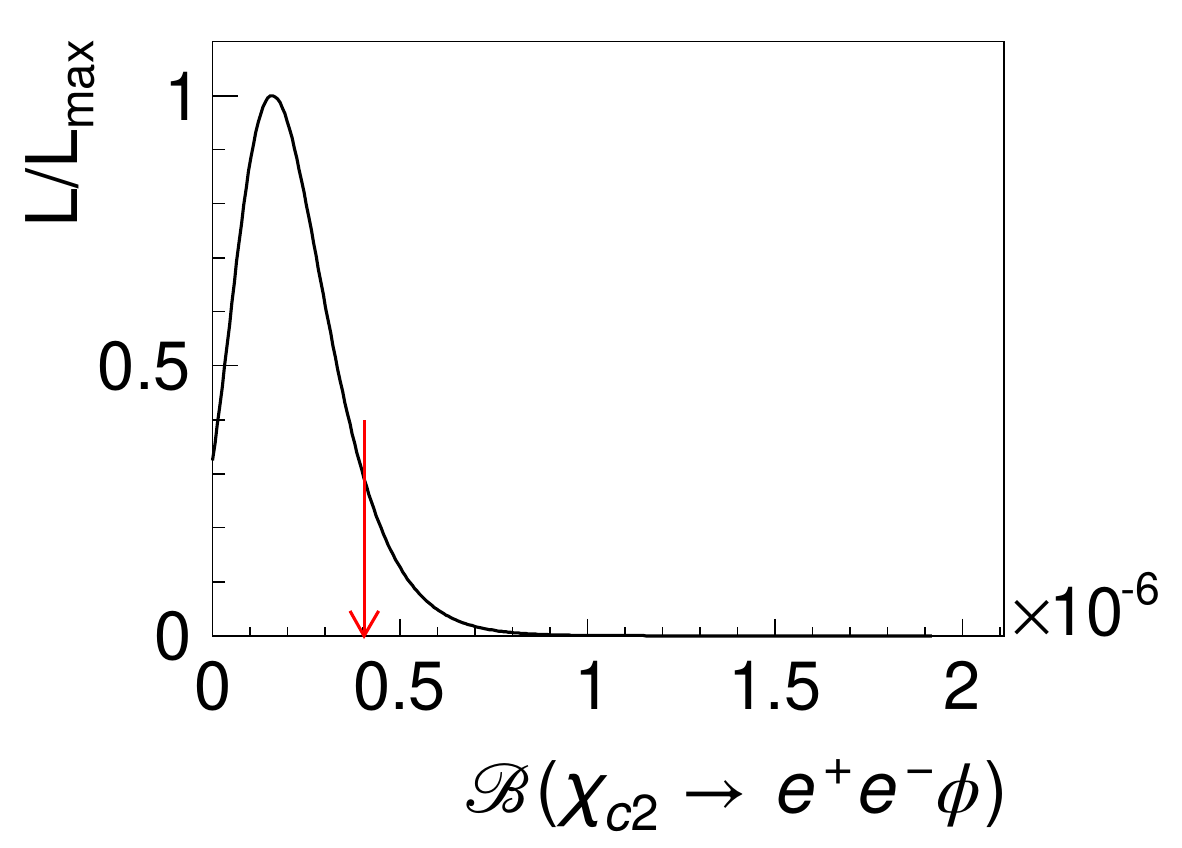}
		\label{}}a 
	\caption{The profile likelihoods of the branching fractions of (a) $\chi_{c0}$, (b) $\chi_{c1}$, and (c) $\chi_{c2}$ decays into $e^+e^-\phi$. The red arrows show the upper limits on individual branching fractions at 90\% C.L..}
    \label{ul}
\end{figure*}
For each signal decay, the upper limit on the number of the signal events, $N^{\rm{UL}}$, at 90\% confidence level~(C.L.) is calculated with an unbound profile likelihood treatment of the systematic uncertainties~\cite{ROLKE2005493}. 

The likelihood function is defined as

\begin{widetext}
\begin{equation}
\begin{aligned}
    \mathcal{L}(\mu, \epsilon_{\rm{sig}}, N_{\rm{bkg}i})= \mathcal{P}(N_{\rm{obs}}^{\rm{count}},\mu\cdot\epsilon_{\rm{sig}}+N_{\rm{bkg}1}+N_{\rm{bkg}2}-N_{\rm{bkg}3})\times\mathcal{G}(\epsilon_{\rm{sig}},\epsilon^{\rm{MC}}_{\rm{sig}},\delta_{\epsilon}\cdot\epsilon^{\rm{MC}}_{\rm{sig}})\times\prod_i \mathcal{P}(N_{\rm{bkg}i}^{\rm{count}},\tau_i\cdot N_{\rm{bkg}i}).
    \end{aligned}\label{eq3}
\end{equation}
\end{widetext}
Here, $N_{\rm{obs}}^{\rm{count}}$ is the number of observed events in the signal region; $\mu$ is the expected number of signal events; $N_{\rm{bkg}1},~N_{\rm{bkg}2}$, and $N_{\rm{bkg}3}$ are the expected numbers of non-$\phi$ resonance background events in the signal region, of the non-$\chi_{cJ}$ events in the signal region, and of the neither $\phi$ nor $\chi_{cJ}$ events in the signal region, respectively. The detection efficiency $\epsilon_{\rm{sig}}$ follows a Gaussian distribution ($\mathcal{G}$) with the mean value $\epsilon^{\rm{MC}} _{\rm{sig}}$ estimated by signal MC samples and the absolute uncertainty $\delta_{\epsilon}\cdot\epsilon^{\rm{MC}}_{\rm{sig}}$, where $\delta_{\epsilon}$ is the relative systematic uncertainty. The numbers of observed events in the sideband region, $N_{\rm{bkg}i}^{\rm{count}}$, follow a Poisson distribution ($\mathcal{P}$) with expected values $\tau_i\cdot N_{\rm{bkg}i}$. Here, $\tau_1,~\tau_2$, and $\tau_3$ are the ratios of the area of the $\phi$ sideband region and the signal region, the area of the $\chi_{cJ}$ sideband region and the signal region, and the area of neither $\phi$ nor $\chi_{cJ}$ sideband region and the signal region, respectively. $N_{\rm{bkg}1}^{\rm{count}},~N_{\rm{bkg}2}^{\rm{count}}$, and $N_{\rm{bkg}3}^{\rm{count}}$ are the number of observed events in the $\phi$ sideband regions, of observed events in the $\chi_{cJ}$ sideband regions, and of observed events neither in the $\phi$ sideband region nor in the $\chi_{cJ}$ sideband region, respectively.

To estimate the upper limit of $\mathcal{B}(\chi_{cJ}\to e^+e^-\phi)$, we calculate the likelihood function of the scanned branching fraction. For a sequence of fixed positive values of $\mathcal{B}$, the likelihood values
are maximized with respect to all other variables. The likelihood distributions are normalized to the maximum likelihood value $\mathcal{L}_{\rm{max}}$. 

The upper limits on the $\mathcal{B}(\chi_{cJ}\to e^+e^-\phi)$ at 90\% C.L. are set as the values when the integral of the profile likelihood curve from zero reaches 90\%.
Considering the influence of the sideband regions choice, we vary the $\phi$ sideband region by $\pm 2.5$ MeV$/c^2$  and the $\chi_{cJ}$ sideband region by $\pm 5$ MeV$/c^2$;  the most conservative upper limit result is considered as the final result. The profile likelihood curve and the final result are shown in FIG.~\ref{ul}.

Finally, the upper limits on $\mathcal{B}(\chi_{cJ}\to e^+e^-\phi)$ are set to $2.4\times10^{-7}$, $6.7\times10^{-7}$, and $4.1\times10^{-7}$, at 90\% C.L. for $\chi_{cJ}\to e^+e^-\phi~(J=0,\,1,\,2)$, respectively. It is worth emphasizing that in the selection of $\chi_{cJ}\to e^+e^-\phi$ events, we require that both $e^+$ and $e^-$ originate from a virtual photon~($\gamma^*$), thus excluding the contributions from $e^+e^-$ pairs originating from the $\phi$ resonance in the upper limit calculations. No other vector resonances are observed in the $e^+e^-$ invariant mass  distribution in the data samples.

\section{SUMMARY}
In summary, by analyzing $(2.712 \pm 0.014)\times10^{9}$ $\psi(3686)$ events collected at $\sqrt{s}$ = 3.686 GeV with the BESIII detector at the BEPCII collider, no significant $\chi_{cJ}\to e^+e^-\phi$ signals are observed. The upper limits on $\mathcal{B}(\chi_{cJ}\to e^+e^-\phi)$ are set to $2.4\times10^{-7}$, $6.7\times10^{-7}$, $4.1\times10^{-7}$, at 90\% C.L. for $\chi_{cJ}\to e^+e^-\phi~(J = 0, 1, 2)$, respectively. These are the first searches of the EM Dalitz decays $\chi_{cJ}\to e^+e^-\phi$. The $\mathcal{B}(\chi_{cJ}\to e^+e^-\phi)$ and TFF can be well measured at future high luminosity facilities, such as a super tau-charm facility~\cite{Achasov:2023gey}.

\input{acknowledgement_2025-02-08}

\bibliography{apssamp}

\end{document}

%% file: authorlist_2025-02-08.tex
\author{
M.~Ablikim$^{1}$\BESIIIorcid{0000-0002-3935-619X},
M.~N.~Achasov$^{4,b}$\BESIIIorcid{0000-0002-9400-8622},
P.~Adlarson$^{77}$\BESIIIorcid{0000-0001-6280-3851},
X.~C.~Ai$^{82}$\BESIIIorcid{0000-0003-3856-2415},
R.~Aliberti$^{36}$\BESIIIorcid{0000-0003-3500-4012},
A.~Amoroso$^{76A,76C}$\BESIIIorcid{0000-0002-3095-8610},
Q.~An$^{73,59,\dagger}$,
Y.~Bai$^{58}$\BESIIIorcid{0000-0001-6593-5665},
O.~Bakina$^{37}$\BESIIIorcid{0009-0005-0719-7461},
Y.~Ban$^{47,g}$\BESIIIorcid{0000-0002-1912-0374}
% ...
\\
\vspace{0.2cm}
(BESIII Collaboration)\\
\vspace{0.2cm} {\it
$^{1}$ Institute of High Energy Physics, Beijing 100049, People's Republic of China\\
% ...
$^{4}$ Budker Institute of Nuclear Physics SB RAS (BINP), Novosibirsk 630090, Russia\\
% ...
$^{36}$ Johannes Gutenberg University of Mainz, Johann-Joachim-Becher-Weg 45, D-55099 Mainz, Germany\\
$^{37}$ Joint Institute for Nuclear Research, 141980 Dubna, Moscow region, Russia\\
% ...
$^{47}$ Peking University, Beijing 100871, People's Republic of China\\
% ...
$^{58}$ Southeast University, Nanjing 211100, People's Republic of China\\
$^{59}$ State Key Laboratory of Particle Detection and Electronics, Beijing 100049, Hefei 230026, People's Republic of China\\
% ...
$^{73}$ University of Science and Technology of China, Hefei 230026, People's Republic of China\\
% ...
$^{76}$ University of Turin and INFN, (A)University of Turin, I-10125, Turin, Italy; (B)University of Eastern Piedmont, I-15121, Alessandria, Italy; (C)INFN, I-10125, Turin, Italy\\
$^{77}$ Uppsala University, Box 516, SE-75120 Uppsala, Sweden\\
% ...
$^{82}$ Zhengzhou University, Zhengzhou 450001, People's Republic of China\\
\vspace{0.2cm}
$^{\dagger}$ Deceased\\
% ...
$^{b}$ Also at the Novosibirsk State University, Novosibirsk, 630090, Russia\\
% ...
$^{g}$ Also at State Key Laboratory of Nuclear Physics and Technology, Peking University, Beijing 100871, People's Republic of China\\
% ...
}
}

%% file: acknowledgement_2025-02-08.tex
\begin{acknowledgments}

The BESIII Collaboration thanks the staff of BEPCII (https://cstr.cn/31109.02.BEPC) and the IHEP computing center for their strong support. This work is supported in part by National Key R\&D Program of China under Contracts Nos. 2025YFA1613900, 2023YFA1606000, 2023YFA1606704; National Natural Science Foundation of China (NSFC) under Contracts Nos. 11635010, 11935015, 11935016, 11935018, 12025502, 12035009, 12035013, 12061131003, 12192260, 12192261, 12192262, 12192263, 12192264, 12192265, 12221005, 12225509, 12235017, 12361141819, 12475093; the Chinese Academy of Sciences (CAS) Large-Scale Scientific Facility Program; CAS under Contract No. YSBR-101; 100 Talents Program of CAS; State Key Laboratory of Nuclear Physics and Technology, Peking University under Contracts No. NPT2025KFY03; The Institute of Nuclear and Particle Physics (INPAC) and Shanghai Key Laboratory for Particle Physics and Cosmology; German Research Foundation DFG under Contract No. FOR5327; Istituto Nazionale di Fisica Nucleare, Italy; Knut and Alice Wallenberg Foundation under Contracts Nos. 2021.0174, 2021.0299; Ministry of Development of Turkey under Contract No. DPT2006K-120470; National Research Foundation of Korea under Contract No. NRF-2022R1A2C1092335; National Science and Technology fund of Mongolia; National Science Research and Innovation Fund (NSRF) via the Program Management Unit for Human Resources \& Institutional Development, Research and Innovation of Thailand under Contract No. B50G670107; Polish National Science Centre under Contract No. 2024/53/B/ST2/00975; Swedish Research Council under Contract No. 2019.04595; U. S. Department of Energy under Contract No. DE-FG02-05ER41374


\end{acknowledgments}